\documentstyle{article}

\def\ben{\begin{equation}}
\def\een{\end{equation}}
\def\bena{\begin{eqnarray}}
\def\eena{\end{eqnarray}}

\begin{document}

\hfuzz=100pt
\title{Holography and the Future Tube} 
\author{G W Gibbons \footnote{permanent address DAMTP, University of Cambridge},
\\
Laboratoire de Physique Th\'eorique de l'Ecole Normale Sup\'erieure \footnote{ Unit\'e Mixte de Recherche de Centre National de la Recherche Scientifique et de l'Ecole Normale Superi\'eure },
\\ 24 Rue Lhomond,
\\ 75231 Paris Cedex 05, 
 \\ France}

\maketitle
\begin{abstract}
The Future Tube $T^+_n$
of $n$-dimensional Minkowski spacetime may be identified with
the reduced phase space $P$, or \lq \lq space of motions " of a particle
moving in $(n+1)$-dimensional Anti-de-Sitter Spacetime.  Both are
isomorphic to a homogeneous
bounded domain in ${\bf C}^{n}$  whose Shilov boundary 
may be identified with $n$-dimensional  conformally compactified
Minkowski spacetime.

\end{abstract}

\section{Introduction}The purpose of this talk is to describe some 
remarkable geometric facts relating
the Future Tube $T^+_n$
of $n$-dimensional Minkowski spacetime  to
the reduced phase space $P$, or \lq \lq space of motions " of a particle
moving in $(n+1)$-dimensional Anti-de-Sitter Spacetime 
with a view to illuminating the Maldacena conjectures \cite{JM}
relating string theory on $AdS_{n+1}$ to Conformal Field theory
on its conformal boundary.

\section{The Future Tube}Slightly confusingly perhaps,
the Future Tube $T^+_n$ is usually defined as those points of complexified
Minkowski spacetime $ {\bf E}^{n-1,1}_{\bf C}$, 
with complex coordinates $z^\mu=x^\mu+i y^\mu$,
such that $y^\mu$ lie inside the  past light cone $C^-_{n}$. That is $y^0 < -\sqrt{y^iy^i}$.
With this convention together with standard conventions of quantum field theory,
a positive frequency function is the boundary value
of  a function which  is holomorphic in the
future tube. The relation between a holomorphic function, for example
one defined 
in some bounded domain $D \subset {\bf C}^n$, and the values of
that function on its Shilov boundary $S$,
which is an $n$ real dimensional submanifold of
the $(2n-1)$ real dimensional boundary $\partial D$  may be said to be \lq \lq holographic \rq \rq in that the information about two real valued
functions of $2n$ real variables is captured by two
real valued functions of $n$ real variables. This is the key idea
behind the application of dispersion relations to quantum field theory
and their use to derive rigourous general results such as the spin-statistics theorem and invariance under CPT.

\section{Bounded Complex Domains}
Homogeneous bounded domains were classified by Cartan \cite{Car}
and their properties are described in detail by Hua \cite{Hua}. 
The case we are interested in corresponds to the Hermitian symmetric space
$D=SO(n,2)/(SO(2)\times SO(n))$. It is referred to by Hua \cite{Hua}
as \lq \lq Lie Sphere Space \rq \rq. In order to see this,
consider the complex light cone  $C^{n+2} _{\bf C} \subset  {\bf C}^{n+2}$ given by 
\ben
(W^{n+1})^2 + (W^{n+2})^2 -(W^i)^2=0 \label{cone}.
\een
Compactified complexified Minkowski spacetime $\overline {{\bf E}_{\bf C}^{n-1,1} } $ consists of complex
light rays passing through the origin. This means that one must identify rays 
$ W^A$ and $\lambda W^A$, where $A=1,\dots, n+2$, $i=1,\dots, n$
and $\lambda \in {\bf C}^\star \equiv {\bf C} \setminus 0$,
that is $\overline {{\bf E}_{\bf C}^{n-1,1}} = C^{n+2} _{\bf C} /{\bf C}^\star$.
Evidently $SO(n-1,2;{\bf R})$ acting in the obvious way on ${\bf C}^{n+1}$
leaves the complex lightcone
$C^{n+2} _{\bf C} $ invariant and commutes with the 
${\bf C}^\star$ action. Thus
the action of $SO(n-1,2;{\bf R})$ descends to $D$.
If we restrict the coordinates $W^A$ to be real, we obtain
the standard construction of $n$-dimensional compactified real Minkowski
spacetime  $ \overline{ {\bf E}^{n-1,1} }$,
as light rays through the origin of ${\bf E}^{n-1,2}$.
It follows that $ \overline {{\bf E}^{n-1,1} } \equiv (S^1\times S^{n-1})/{\bf Z}_2$, The finite points in Minkowski spacetime, i.e. those not contained
in what Penrose refers to as $\cal I$ may be obtained by intersecting
the light cone with a null hyperplane which does not pass through
the origin.

To get to the description given by Hua \cite{Hua}, introduce coordinates 
parameterising (most of) the light cone by
$u,w^i$ by
\ben
W^{n+1}-iW^{n+2} ={ 1\over u},
\een
\ben
W^{n+1}+iW^{n+2}={w^i w^i \over {u}},
\een
and 
\ben
W^i= {w^i\over u},
\een
If $w\in {\bf C}^{n}$ is 
a complex $(n-1)$ column vector and $w^2=w^tw$ and $|w|^2=w^\dagger w$
then the domain $D$  defined by \cite{Hua}
\ben
1-|w|^2\ge \sqrt{ |w|^4-|w^2|^2}.
\een
The topological boundary is given by the real equation:
\ben
1-2|w|^2 +|w^2|^2=0.
\een
On the other hand, the  Shilov boundary $S \subset \partial D$
is determined by the property that
the maximum modulus of any holomorphic function on $P$ is attained
on $S$. Consider, for example,
the holomorphic function $w$. 
It attains its maximum modulus
when $w=\exp(i\tau) {\bf n}$, where $\bf n$ is a real
unit $(n-1)$ vector. Thus $S$ is given by $S^1\times S^{n-1}/{\bf Z}_2$.
If we take $W^i$ to be the  real unit $n$-vector $\bf n$
 and $u= \exp(-i \tau)$
we see that $S$ and  $ \overline {{\bf E}^{n-1,1} } $ are one and the same thing.

\section{The geodesic Flow of $AdS_{n+1}$}

Now let us turn to $(n+1)$-dimensional
Anti-de-Sitter spacetime .
One possible approach to quantising a relativistic particle
moving in $AdS_{n+1}n$, might be  to 
look at the relativistic phase space
and then pass to the constrained space  and  to ``quantise" it.
Recall that, in general, the relativistic phase space of
a spacetime $M$ is the cotangent bundle $T^\star M$
with coordinates $\{x^\mu, p_\mu\}$, canonical one-form $A=p_\mu dx^\mu$
and symplectic form
\ben
\omega = dp_\mu \wedge dx^\mu.
\een
The geodesic flow is generated by the covariant
Hamiltonian
\ben
{\cal H}={ 1\over 2}g^{\mu \nu} p_\mu p_\nu.
\een
The flow for a timelike geodesic, corresponding to a
particle of mass $m$ lies on the level sets, call them $\Gamma$,
given by
\ben
{\cal H}= -{ 1\over 2} m^2.
\een
The restriction of the canonical one-form $A$ to the level sets 
$\Gamma$ endows them with a contact structure,
in other words, the restriction of $dA$ has rank $2n$ and it's 
one-dimensional kernel is directed along the geodesic flow. Thus,
locally at least, one may 
pass to the reduced phase space $P=\Gamma/G_1 $
where $G_1$ is the one-parameter group generated by the 
covariant Hamiltonian
$\cal H$, by a ``Marsden-Weinstein reduction".
Geometrically, the group $G_1$ 
takes points and their cotangent vectors along the world lines of the 
timelike geodesics.

The reduced $2n$-dimensional
phase space $P$ is naturally a symplectic manifold.
Moreover the isometries of $M$  act 
by canonical transformations, i.e. by symplectomorphisms,
on $P$ taking timelike geodesics to timelike geodesics.
Each Killing vector field $K_a^\mu(x)$ on $M$ determines
a ``moment map" $\mu_a(x,p) = K^\mu_a p_\mu$ on $T^\star M$ which Poisson
commutes with the covariant Hamiltonian $\cal H$. 
They thus descend to reduced phase space $P$
where their Poisson algebra is is the same as the Lie algebra  of
of the isometry group. Thus, for example, if $n=2$
the Poisson algebra is $sl(2;{\bf R})$. This fact
is behind
the connection between  black holes, conformal mechanics, and 
Calogero models
discussed in \cite{GTII}.

We are interested in the quantum theory rather
than the classical geodesic motion and so it is appealing
to attempt to implement the 
geometric quantization programme by ``quantising " $P$.
A point of particular interest would then be to compare it with
a more conventional approach based on quantum field theory 
in a fixed background.
In the general case this appears to be difficult because one does not have a good
understanding of the space of timelike geodesics $P$.   However
in the present case of
$AdS_{n+1}$, the space $P$ may be described rather explicitly. 
It is a homogeneous K\"ahler manifold which is isomorphic to the future tube $T^+_{n}$
of $n$-dimensional Minkowski spacetime. 
Because it is a K\"ahler manifold
one may adopt a holomorphic polarisation. The resulting ``quantization" is
the same as that considered by Berezin and others
many years ago (see e.g. \cite{Ber1,Ber2, Ber3}). A more physical
description is in terms of coherent states.

To see the relation between $P$ and $T^+_n$ 
explicitly it is convenient to regard   $AdS_{n+1}$ as the real
quadric in ${\bf E}^{n,2}$ given by
 \ben
(W^{n+1})^2 + (W^{n+2})^2 -(W^i)^2=1 \label{ads}.
\een
It is clear by comparing (\ref{cone})
and (\ref{ads}) that the light cone $C_{n+2}$ and the quadric
$AdS_{n+1}$ can only touch at infinity, which explains why
the conformal boundary of $AdS_{n+1}/{\bf Z}_2$ is the same as
compactified Minkowski spacetime 
$\overline {{\bf E}^{n-1,1}} \equiv (S^1 \times S^{n-1}) /{\bf Z}_2$,
where ${\bf Z}_2$ is the antipodal map $\bf Z_2: W^A \rightarrow -W^A$.

Using this representation of $AdS_{n+1}$
, one easily sees that every timelike geodesic is 
equivalent to every other one under
an $SO(n-1,2)$ transformation. They  may all be obtained as the intersection
of some
 totally timelike
2-plane passing through the origin of
of the embedding space ${\bf E}^{n,2}$ with
the $AdS_{n+1}$ quadric. The space $P$
of such two planes may thus be identifed
with the space of geodesics. It is 
a homogeneous space of the isometry  group,
in fact the Grassmannian $SO(n,2) /(SO(2) \times SO(n))$.
Note that, as one expects, the dimension of $P$ is $2n$. The denominator
of the coset is the maximal compact subgroup of $SO(n,2)$. Two factors
correspond to timelike rotations in the  timelike 2-plane and rotations of
the normal space respectively. The former may be identified with the one
parameter group $G_1$ generated by the covariant Hamiltonian $\cal H$. Thus
the level set $\Gamma$ is the coset space $SO(n,2) / SO(n)$.

\section{Quantisation}

Given a manifold $X$ with coordinates $x$ and a measure $\mu$, an over complete set of coherent
states is a set of vectors $\{ |x \rangle \} $ in some quantum mechanical
Hilbert space $H_{\rm qm}$ providing a resolution of the identity
\ben
\int _X\mu   |x \rangle \langle x| =\hat 1.
\een
Usually one takes $X$ to be a homogeneous space of some
Lie group. However that is not essential
for the general concept.
A Hermitian operator $\hat A$ on  $H_ {\rm qm}$ is associated with a real
function $A(x)$ via the relation
\ben
\hat A = \int_X  A(x) \mu   |x\rangle \langle x| =\hat 1.
\een
The commutator algebra of a set of operators on   $H_{\rm qm}$ then gives rise
to an algebra $\cal A$
on the associated functions. If $X$ is
a symplectic manifold one expects, at least in the limit that 
Planck's constant  is small, or more strictly speaking in the limit
of large action,
that the  algebra $\cal A$
reflects the Poisson
algebra on the functions on $X$. For the special
case of non-compact K\"ahler manifold of dimension $2n$,
K\"ahler form $\omega$ and with coordinates $w^i$
and  metric
\ben
g_{i\overline j} ={ \partial ^2 F \over \partial w^i \partial 
{\overline w^j} }.
 \een
where $F$ is the K\"ahler potential it was proposed by Berezin \cite{Ber1,Ber2,Ber3}
that
one choose for $H_{\rm qm}$ the space of holomorphic functions
with inner product
\ben
\langle g(w) |f(w) \rangle =  \int_X \overline {g(w)} f(z) {\omega ^n
\over n!} \exp (- {1 \over h}F)
\een
One must now choose $h$ so that one gets a non-trivial Hilbert
space $H_{\rm qm}$. Given that one may proceed to
represent the isometries of $X$ on $H_{\rm qm}$ and to introduce other
operators and investigate their algebras.
Note that the metric on $H_{\rm qm}$ will in general
depend upon $h$. The quantity $h$ is referred to in this context as Planck's
constant, although physically, for dimensional reasons,
that is not really accurate. If it happens that $X$ is Einstein K\"ahler
with negative scalar curvature then the Monge Amp\`ere equation
tells us that 
\ben
{\rm det} g_{i \overline j}= \exp (-{\Lambda F})
\een
Thus, to get convergence one wants $h \le {1 \over |\Lambda |}$.
An upper bound for Planck's constant seems very puzzling from
a physical point of view but it has a simple explanation.

Consider the circle bundle $S^1 \rightarrow E \rightarrow X$
over the K\"ahler manifold $X$ with a connection
whose curvature $F=dA$ is a multiple, $e$,
of the symplectic form $\omega$. 
Thus $F=e\omega$ and  one may think of $e$ as the product of
the electric charge and  the strength of magnetic field.
The covariant derivative is 
$\overline {\cal D} =\overline \partial + e \partial _{\overline w}F$.
Thus if $\psi=e^{-eF} f(w)$, where $f(w)$ is holomorphic, then
\ben 
\overline {\cal D} \psi =0.
\een
Now the space of spinors on a K\"ahler manifold may be identified with
the space of differential forms of type (0,p). The Dirac operator
corresponds to the operator $\sqrt 2 ({\overline \partial} + {\overline
\partial}^\dagger)$. Minimally coupling the Dirac operator to the
K\"ahler connection corresponds geometrically to taking the tensor product
of the space of spinors with a power of the canonical bundle. The first power
gives the canonical $Spin^c$ structure on $X$.
If $X$ is assumed to have trivial second co-homology
we may take any (not even rational power).
The Dirac operator minimally coupled to $F$
corresponds to $\sqrt 2 (\overline {\cal D} + \overline {\cal D}^\dagger )$.
It follows that by setting $2e=-{ 1\over h}$ We may identify $H_{\rm qm}$ with
the space of zero modes of the minimally coupled Dirac operator on $X$.
Now typically $X$ has negative curvature, and so only if the charge
on the spinors is sufficiently large, and of the correct sign, will
there be a big space of, or indeed any, zero modes.

The above theory was rather general. We now restrict to
the case of
a bounded complex domain $D$ in ${\bf C}^n$. We begin by describing its 
K\"ahler structure.
Associated with $D$ is  
the  Hilbert space ${\rm Hol}(D)$ of square integrable holomorphic
functions. 
If $\{\phi_s \}$, $s=1,\dots, $ is an orthonormal basis for  ${\rm Hol}(D)$
the Bergman Kernel $K(w,\overline v)$ is defined by
\ben
K(w,\overline v)= \sum \phi_s(w) \overline {\phi_s(v)}.
\een
The Bergman Kernel gives rise to a K\"ahler potential
 $F(w, \overline w)=\log K(w,\overline w)$ in terms of which
 the Bergman metric on $D$ is given by
\ben
g_{i\overline j} ={ \partial ^2 F \over \partial w^i \partial 
{\overline w^j} }.
 \een
Geometrically the basis $\{\phi_s \}$ gives a map of $D$ into
${\bf C }{\bf P}^{\infty}$ and $ g_{i\overline j}$ is the pull-back of the Fubini-Study
metric. Now although $D$ has finite Euclidean volume, because
 $K$ and $F$ typically diverge at the boundary $\partial D$
the volume of $D$ with respect to the K\"ahler metric $ g_{i\overline j}$
diverges. For example, in our case Hua \cite{Hua}
has calculated $K(w,\overline w)$ and
finds
\ben
K(w,\overline w)= { 1\over V_n} { 1\over (1+ |w^2|^2 -2 |w|^2 )^{n}},
\een
where $V_n= { \pi ^n \over 2^{n-1} n!}$ is the Euclidean volume of $D$.

\section{Cheng-Mok-Yau-Anti-de-Sitter Spacetimes}

The theory described above does not take into account gravity.
Of course  supergravity methods provide a way of doing that.
They would lead to replacing $AdS_n$ by some other
solution of the supergravity equations of motion.
For example with some other Einstein Space. 
A great deal of work has already appeared going in this direction.
One new possibility will be described in this section.

However it is also worth asking how the space $P$
might be generalised. The relationship between these two
generalisations is then of interest. 
This will be dercibed in the next section.

If $n$ is odd, $n=2m+1$ say, then $AdS_{2m+1}$ may be regarded as 
a circle bundle over complex hyperbolic space $H^m_{\bf C}$ \cite{GP}.
Clearly, using complex coordinates $Z^A$, $A=1,\dots,m+1$ in 
${\bf R}^{2m+2} \equiv {\bf C}^{m+1}$, $AdS_{2m+1}$ is given by

\ben
|Z^1|^2 - \dots - |Z^{m+1}|^2=1.
\een

We may now fibre by  the  $U(1)$ action $Z^A \rightarrow e^{it} Z^A$.
The orbits are 
{\sl closed timelike curves} in $AdS_{2m+1}$. The base space $B$ has a Riemannian, i.e. positive definite, metric.
In fact $ B= SU(m,1)/U(m)\equiv H^m_{\bf C}$
is the unit ball in ${\bf C}^m$. This is another bounded
complex domain in ${\bf C}^m$.  The metric on the base space 
is precisely its  Bergman metric.
In fact  $H^m_{\bf C}$ with its Bergman metric is 
ithe symmetric space dual
of ${\bf C} {\bf P}^m$  with its 
Fubini-Study metric. The construction we have just described
is the symmetric space dual of the usual  Hopf fibration.

The metric is
\ben
ds^2= -(dt+A_adx^b)^2 + g_{ab}dx^a dx ^b
\een
where $a, b=1,2,\dots ,2m$, $g_{ab}$ is the Einstein-K\"ahler
metric and $dA$ is the K\"ahler form.
In traditional relativist's
language, $AdS_{2m+1}$ has been 
exhibited as an  ultra-stationary metric
(i.e.  one with constant Newtonian potential
$U={ 1\over 2} \log (-g_{00})$.
 The Sagnac or gravito-magnetic connection, governing frame-dragging
effects corresponds precisely to the connection of the standard circle
bundle over the K\"ahler base space. Its curvature is the K\"ahler form.

Now it is easy to check that one may replace the Bergman manifold 
$\{B,g_{ab}\}$ with any other $2m$ dimensional
Einstein-K\"ahler manifold with negative scalar curvature and obtain a
$(2m+1)$-dimensional Lorentzian Einstein manifolds admitting Killing spinors
in this way. According to Cheng and Yau \cite{CY}
and Mok and Yau \cite{MY}, there is a rich 
supply of complete
Einstein K\"ahler metrics on complex domains. Their investigation
might well prove to be fruitful. The conformal boundary
of these spacetimes is clearly related to the boundary of the complex domain,
but the precise relationship is not at all clear. Apart from their possible
applications to the Maldecena conjecture, these spacetimes may provide 
a useful arena, albeit in higher dimensions,
 for investigating the effects of closed timelike curves in general relativity.

\section{Adapted Complex Structures}

We have seen that the space $P$  of timelike
geodesics in $AdS_n$ carries an Einstein
K\"ahler structure. In fact
the entire cotangent bundle $T^\star AdS_n$ admits a Ricci-flat pseudo
K\"ahler metric, i.e. one with signature $(2n-2,2)$.
The existence of this Ricci-flat pseudo-K\"ahler metric may be obtained
by analytically continuing Stenzels's \cite{Sten} positive definite
Ricci-flat K\"ahler metric on 
the cotangent bundle of the standard $n$-sphere, $T^ \star S^n$ \cite{Sten}.
The simplest case is when $n=2$. Stenzel's metric is then
the  Eguchi-Hanson metric which may be analytically continued to give a 
``Kleinian" metric of signature $(2,1)$ on $T^\star AdS_2$. In fact
the cotangent bundle of {\sl any} Riemannian manifold may be endowed with
a canonical complex structure and a variety of K\"ahler metrics.
We will describe this construction shortly and formulate
a  version for Lorentzian manifolds. Before doing so we descibe
Stenzel's construction.

The cotangent bundle of the $n$-sphere  $T^\star S^{n}$
may be identified with an affine quadric in ${\bf C}^{n+1}$.
This may be seen as follows: $T^\star S^{n}$ consist of a pair of real $(n+1)$
vectors $X^A$ and $P^A$ such that
\ben
X^1 X^1 +X^2 X^2+\dots + X^{n+1}X^{n+1}=1,
\een
\ben
X^1 P^1 + X^2 P^2 +\dots + X^{n+1} P^{n+1}=0.
\een
If $P=\sqrt{ P^1P^1+ P^2P^2 +\dots +P^{n+1}P^{n+1}}$ 
one may map $T^ \star S^{n}$ into the affine quadric 
\ben
(Z^1)^2  + (Z^2)^2+\dots +(Z^{n+1})^2=1 \label{quadric}
\een
setting
\ben
Z^A=A^A+iB^A=\cosh(P) X^A  +i{ \sinh(P) \over P}P^A. 
\een

Stenzel then seeks a K\"ahler potential depending only on the restriction
to the quadric (\ref{quadric}) of the function
\ben
\tau= |Z^1|^2+|Z^2|^2 +\dots +|Z^{n+1}|^2.
\een
The Monge-Amp\`ere equation now
reduces to any ordinary differential equation.

In the case of $AdS_{p+2}$ we may proceed as follows.
The bundle of future directed timelike vectors 
in $AdS_{p+2}$, $T^+AdS_{p+2}$  consists of  pairs of timelike vectors
$X^A$, $P^A$ in ${\bf E}^{p+1,2}$ such that
\ben
X^A X^B \eta_{AB}=-1
\een
and
\ben
X^A P^B\eta_{AB}=0,
\een
with $P^A$ future directed and $\eta_{AB}={\rm diag}(-1,-1,+1,\dots ,+1)$ the
metric. We define $P=\sqrt{ {-P^AP^B\eta_{AB}}}$
and 
\ben
Z^A= \cosh(P) X^A +i  {\sinh(P) \over P} P^A
\een
which maps $T^+AdS_{p+2}$ to the affine  quadric 
\ben
Z^A Z^B \eta _{AB}=-1.
\een

One then seeks a K\"ahler potential depending only on the restriction
to the quadric (\ref{quadric}) of the function
\ben
\tau= |Z^{0} |^2+|Z^{p+2} |^2 -|Z^1|^2 - \dots -|Z^{p+1}|^2.
\een
The Monge-Amp\`ere equation again
reduces to an ordinary differential equation.

The canonical complex structure on $T^\star M$ is defined as follows.
Let $\sigma \rightarrow \bigl (x^\mu(\sigma), p_\mu=g_{\mu \nu} { d x^\nu \over d \sigma}(\sigma )  \bigr )$
 be a solution of Hamilton's equations for a geodesic in $T^\star M$.
Then in any complex chart $w^i$, $i=1,\dots,n$ one demands that 
for all geodesics the
map ${\bf C} \rightarrow T^\star M$ given by $\sigma +i \tau
\rightarrow \bigl (x^\mu(\sigma), \tau p_\mu \bigr ) $ is holomorphic. 
The ``energy " ${\cal H} = { 1\over 2} g^{\mu \nu} p_\mu p_\nu$
is a real valued positive function on $T^\star M$ which vanishes
only on $M$. It is plurisubharmonic with respect to the complex structure
i.e. the hermitian metric
\ben
{\partial ^2 {\cal H} \over \partial w^i \partial \overline w ^j }
\een
is positive definite, and $\sqrt{\cal H}$ satisfies the homogeneous
Monge-Amp\`ere equation
\ben
{\rm det} \bigl ( {\partial ^2 \sqrt {\cal H} \over \partial w^i \partial \overline w ^ j} \bigr )=0.
\een

In the case of Stenzel's construction one has ${\cal H}={ 1\over 2} P^2
= {1 \over 8} (\cosh ^{-1}(\tau) )^2$.

\section{The Physical Dimension}

The analysis above works for general dimension.
However the case $n=4$ is special since $SO(4,2) \equiv SU(2,2)/{\bf Z}_2$.
One may identify points in real four-dimensional
Minkowski spacetime ${\bf E}^{3,1}$ with two by two Hermitian matrices
$z=x^0+{\bf x} \cdot {\bf \sigma}$.  The future tube $T^+_4$ then
corresponds to complex matrices $x=z^0+{\bf z}\cdot {\bf \sigma}$ whose
imaginary part is positive definite. The Cayley map  \ben z \rightarrow w=
(z-i)(z+i)^{-1} \een  maps this into the bounded holomorphic domain  in
${\bf C}^4$ consisting of the  space  of two by two
complex matrices $w$ satisfying  \ben 1-ww^\dagger >0. \een

For more details the reader is directed to \cite{U1}. 
For this approach to the compactification of Minkowski spacetime see also 
\cite{U2,GSt}

\section{de-Sitter}

The aim of the talk has been to describe some intriguing relations
between the future tube and the covariant phase space of Anti-de-Sitter spacetime
which seem to lie at the heart of holography. One may ask: what about de-Sitter spacetime?
As one might expect, everything goes wrong.
One  problem is that one does not get a positive  definite metric on covariant phase space.
This makes for difficulties in the usual approach to geometric quantisation.
Thie lack of a positive metric is more or less clear because the timelike geodesics are orbits
of the non-compact group $SO(1,1)$. This last fact is also closely related to thermal
radiation seen by geodesics observers  in de-Sitter spacetime. A closely related problem 
concerns the lack of a positive energy generator of $SO(n,1)$ and the consequent impossibility
of de-Sitter supersymmetry.

\section{Final Remarks}
After the talk I became aware of \cite{GB} where a geometric quantisation
approach to Anti-de-Sitter spacetime is described. I was also told of some work
in axiomatic quantum field theory \cite{R1} \cite{R2} which appears to  have some relation to what
I have desribed above.

\section{Acknowledgement}
I should like to thank the organizers for the opportunity to speak at 
such a stimuating conference.

\end{document}